\documentclass[11pt]{article}
\usepackage[utf8]{inputenc}
\pdfoutput=1
\usepackage{amssymb,amsmath,mathrsfs,enumerate}
\usepackage{graphicx,rotate,multicol}
\usepackage{float}
\usepackage{tocloft}
\usepackage{subfig}
\usepackage[margin=10pt,labelfont=bf]{caption}
\usepackage{cite}
\usepackage{soul}
\usepackage[normalem]{ulem}
\usepackage[colorlinks=true,
            linkcolor=blue,
            urlcolor=blue,
            citecolor=teal]{hyperref}
\usepackage{multirow}

\makeatletter
\let\Hy@linktoc\Hy@linktoc@page
\makeatother

\usepackage{color}
\definecolor{ourcolor}{rgb}{0.7, 0.25, 0.05}
\usepackage{tikz,braket}
\long\def\rpl#1!!#2!!{\textcolor{red}{#1} \textcolor{blue}{#2}}

\let\bar=\overline

\def \order(#1){{\mathcal O} \left(#1 \right)}

\evensidemargin=\oddsidemargin
\allowdisplaybreaks

\title{\color{black}{\bf Solar constraints on captured electrophilic dark matter}}

\author {\bf Debajit Bose,$^{a,}$\footnote{debajitbose550@gmail.com} 
\hspace{4pt}  Tarak Nath Maity$^{b,}$\footnote{tarak.maity.physics@gmail.com}
\hspace{4pt}  and Tirtha Sankar Ray$^{a,}$\footnote{tirthasankar.ray@gmail.com} 
\\[10pt]
\small\em $^a$Department of Physics, Indian Institute of Technology Kharagpur, Kharagpur 721302, India\\
\small\em $^b$Centre  for  High  Energy  Physics,  Indian  Institute  of  Science,  Bangalore  560012,  India
}

\date{}

\begin{document}

\maketitle

\begin{abstract}
Dark matter captured by interaction with  electrons inside the Sun may annihilate via long-lived mediator to produce observable gamma ray signals. We utilize solar gamma ray flux measurements from the Fermi Large Area Telescope and High Altitude Water Cherenkov observatory to put bounds on the dark matter electron scattering cross-section. We find that our limits are four to six orders of magnitude stronger than the existing limits for dark matter masses ranging between GeV to PeV scale.
\end{abstract}

%\newpage

\vspace{12 mm}

\hrule \hrule
\tableofcontents
\vskip 10pt
\hrule \hrule 

%%%%%%%%%%%%%%%%%%%%%%%%%%%%%%%%%%%%%%%%%%%%%%%%%%%%
%%%%%%%%%%%%%%%%%%%%%%%%%%%%%%%%%%%%%%%%%%%%%%%%%%%%

\section{Introduction}
\label{sec:intro}

The gravitational wells of celestial bodies can act as local reservoirs for ambient dark matter (DM),  making them attractive location in the sky to search for  it. Typically the relic distributions of particulate DM may be gravitationally focused by the massive stars and subsequently undergo scattering with the stellar constituents dissipating energy in the process. If the resultant energy of the DM particle is below the escape energy, it gets captured within the astrophysical body \cite{Press:1985ug,Gould:1987ju,Gould:1987ir}. Over time a density of captured DM may build up in these celestial objects that are considerably higher than the typical halo distribution making them DM hotspots in the sky. 

Evidently the Sun remains the most significant and by far the most experimentally scrutinized star in the sky making it a sensitive tool to study properties of captured DM \cite{PhysRevLett.55.257,PhysRevD.33.2079,Jungman:1994jr,IceCube:2012ugg,Baratella:2013fya,Danninger:2014xza,Super-Kamiokande:2015xms,Bell:2011sn,Feng:2016ijc,Fornengo:2017lax,Smolinsky:2017fvb,Batell:2009zp,Leane:2017vag,Arina:2017sng,HAWC:2018szf,Niblaeus:2019gjk,Xu:2021glr,Bell:2021pyy,Tonnis:2021krs,Bell:2021esh,Zakeri:2021cur}. While scattering with solar nucleons is the major capture mechanism in the Sun \cite{PhysRevLett.55.257,PhysRevD.33.2079,Jungman:1994jr,Batell:2009zp,Bell:2011sn,IceCube:2012ugg,Baratella:2013fya,Danninger:2014xza,Super-Kamiokande:2015xms,Feng:2016ijc,Fornengo:2017lax,Smolinsky:2017fvb,Leane:2017vag,Arina:2017sng,HAWC:2018szf,Niblaeus:2019gjk,Xu:2021glr,Bell:2021pyy,Tonnis:2021krs,Bell:2021esh,Zakeri:2021cur}. Electrophilic DM particles can instead scatter off solar electrons and get trapped. Such electrophilic DM captured in the Sun can annihilate and produce detectable annihilation signatures owing to the local high density. Neutrino signals from these annihilations have been considered in \cite{Kopp:2009et,Garani:2017jcj}. If the DM particles annihilate via long-lived mediators \cite{Bell:2011sn,Feng:2016ijc,Smolinsky:2017fvb,Batell:2009zp,Leane:2017vag,Arina:2017sng,HAWC:2018szf,Niblaeus:2019gjk,Bell:2021pyy,Leane:2021ihh,Bose:2021yhz,Zakeri:2021cur}, the decay of these escaped mediators can produce detectable gamma ray signatures. In this letter, we demonstrate that the Fermi Large Area Telescope (Fermi-LAT) \cite{Fermi-LAT:2011nwz,Ng:2015gya,Tang:2018wqp} and High Altitude Water Cherenkov (HAWC) observatory \cite{HAWC:2018rpf} are sensitive to such a photon flux. By comparing the solar gamma ray data from Fermi-LAT and HAWC with such a photon flux, we for the first time put constraints on DM-electron scattering cross-section at $\mathcal{O}(10^{-43} - 10^{-40}) \rm \, cm^2$ for DM mass ranging from a few $\rm GeV$ to $\rm 1 \, PeV$. These limits are $\sim 4-6$ orders of magnitude stronger than the current bounds from the other considerations \cite{XENON:2019gfn,Kopp:2009et}.

%%%%%%%%%%%%%%%%%%%%%%%%%%%%%%%%%%%%%%%%%%%%%%%%%%%%
%%%%%%%%%%%%%%%%%%%%%%%%%%%%%%%%%%%%%%%%%%%%%%%%%%%%

%%%%%%%%%%%%%%%%%%%%%%%%%%%%%%%%%%%%%%%%%%%%%%%%%%%%
%%%%%%%%%%%%%%%%%%%%%%%%%%%%%%%%%%%%%%%%%%%%%%%%%%%%

\section{Solar DM capture via electron scattering}
\label{cap}

The gravitationally focused DM particles can scatter with the electrons inside the Sun and get trapped. The capture rate of DM particles is given by \cite{Garani:2017jcj,Liang:2018cjn}
\begin{eqnarray}
\label{eq:caprate}
C_{\odot} = \left( \frac{\rho_{\chi}}{m_{\chi}} \right) \int_0^{R_\odot} dr \, 4 \, \pi \, r^2 \int_0^{u_{\rm esc}} du_{\chi} \frac{f(u_{\chi})}{u_{\chi}} \, w(r) \, \int_0^{v_{\rm esc}(r)} dv \, \, \Omega^{-} \left(w(r) \rightarrow v \right),
\end{eqnarray}
where $\rho_{\chi}$ is the local DM density, $f(u_{\chi})$ is the velocity distribution profile of DM, $m_{\chi}$ is mass of the DM particle and $R_{\odot}$ is the radius of the Sun, $w(r)=\sqrt{u_{\chi}^2 + v_{\rm esc}^2(r)}$ is the velocity of a gravitationally focused DM particle at a distance r from the center of the Sun, $v_{\rm esc}(r)$ being the escape velocity at that location. The galactic escape velocity of DM is given by  $u_{\rm esc} = 528\,$ km/s \cite{Deason2019TheLH,Maity:2020wic}. The differential rate of DM-electron scattering that can trigger a velocity change from $w(r)$ to $v$ can be expressed as
\begin{eqnarray}
\Omega^{-} \left(w(r) \rightarrow v \right) = \frac{2}{\sqrt{\pi}} \frac{(m_{\chi}+m_e)^2}{4 m_{\chi} m_e} \frac{v}{w(r)} \, n_e(r) \, \sigma_{\chi e} \left[ \int_{-\alpha_{-}}^{\alpha_{+}} dy \, e^{-y^2} + 
e^{\frac{m_{\chi} \left(w(r)^2-v^2 \right)}{2 \, T_{\odot}(r)}} \int_{-\beta_{-}}^{\beta_{+}} dy \, e^{-y^2} \right],
\end{eqnarray}
where $m_e$ is the electron mass and $\sigma_{\chi e}$ is the velocity independent DM-electron scattering cross-section. We have utilized the definition \cite{Gould:1987ju}
\begin{eqnarray}
\begin{aligned}
  \alpha_{\pm} = \sqrt{\frac{m_e}{2 \, T_{\odot}(r)}} \, \bigg( \mu_+ \, v \pm \mu_- \, w(r) \bigg), \\
\end{aligned}
\end{eqnarray}
where $\mu_{\pm} = \frac{m_{\chi}}{m_e} \pm 1$, $n_e(r)$ and $T_{\odot}(r)$ are the electron number density and temperature profile of the Sun \cite{Vinyoles:2016djt}. A similar expression can be obtained for $\beta_{\pm} = \alpha_{\pm}(\mu_{\pm} \rightarrow \mu_{\mp})$. As the Sun is moving with a velocity $v_{\odot} = 220 \, {\rm km/s}$ in the galactic halo of DM, the Maxwell-Boltzmann velocity distribution of DM in the rest frame of the Sun is given by \cite{Garani:2017jcj}
\begin{eqnarray}
f(u_{\chi}) = \sqrt{\frac{3}{2 \, \pi}} \frac{u_{\chi}}{v_{\odot} v_d} \left[ e^{-\frac{3 (u_{\chi} - v_{\odot})^2}{2 v_d^2}} - e^{-\frac{3 (u_{\chi} + v_{\odot})^2}{2 v_d^2}} \right],
\end{eqnarray}
where the DM velocity dispersion ($v_d$) is assumed to be $270 \, {\rm km/s}$. Electrons, being light particles, attain a significant thermal velocity within the solar interior which makes the inclusion of solar temperature profile crucial in determining the capture rate \cite{Garani:2017jcj}. For the DM mass range we have explored in the work, the capture rate exhibits a secular decline with increasing mass. 

Evidently a tree level coupling between the DM and the electron can model dependently generate radiative coupling between the DM and nucleon \cite{Kopp:2009et,Bell:2019pyc}. This crucially depends on the lorentz structures of the tree level coupling and can be suppressed for specific choices \cite{Kopp:2009et}. These loop induced couplings would further enhance the capture rate making the limits on DM-electron couplings more stringent. We neglect these model dependent loop induced DM-nucleon couplings and consider conservative limits originating from pure DM-electron interactions.

%%%%%%%%%%%%%%%%%%%%%%%%%%%%%%%%%%%%%%%%%%%%%%%%%%%%
%%%%%%%%%%%%%%%%%%%%%%%%%%%%%%%%%%%%%%%%%%%%%%%%%%%%

%%%%%%%%%%%%%%%%%%%%%%%%%%%%%%%%%%%%%%%%%%%%%%%%%%%%
%%%%%%%%%%%%%%%%%%%%%%%%%%%%%%%%%%%%%%%%%%%%%%%%%%%%

\section{Gamma ray flux}
\label{gamma}

\begin{figure}[t]
\begin{center}
\includegraphics[scale=0.2]{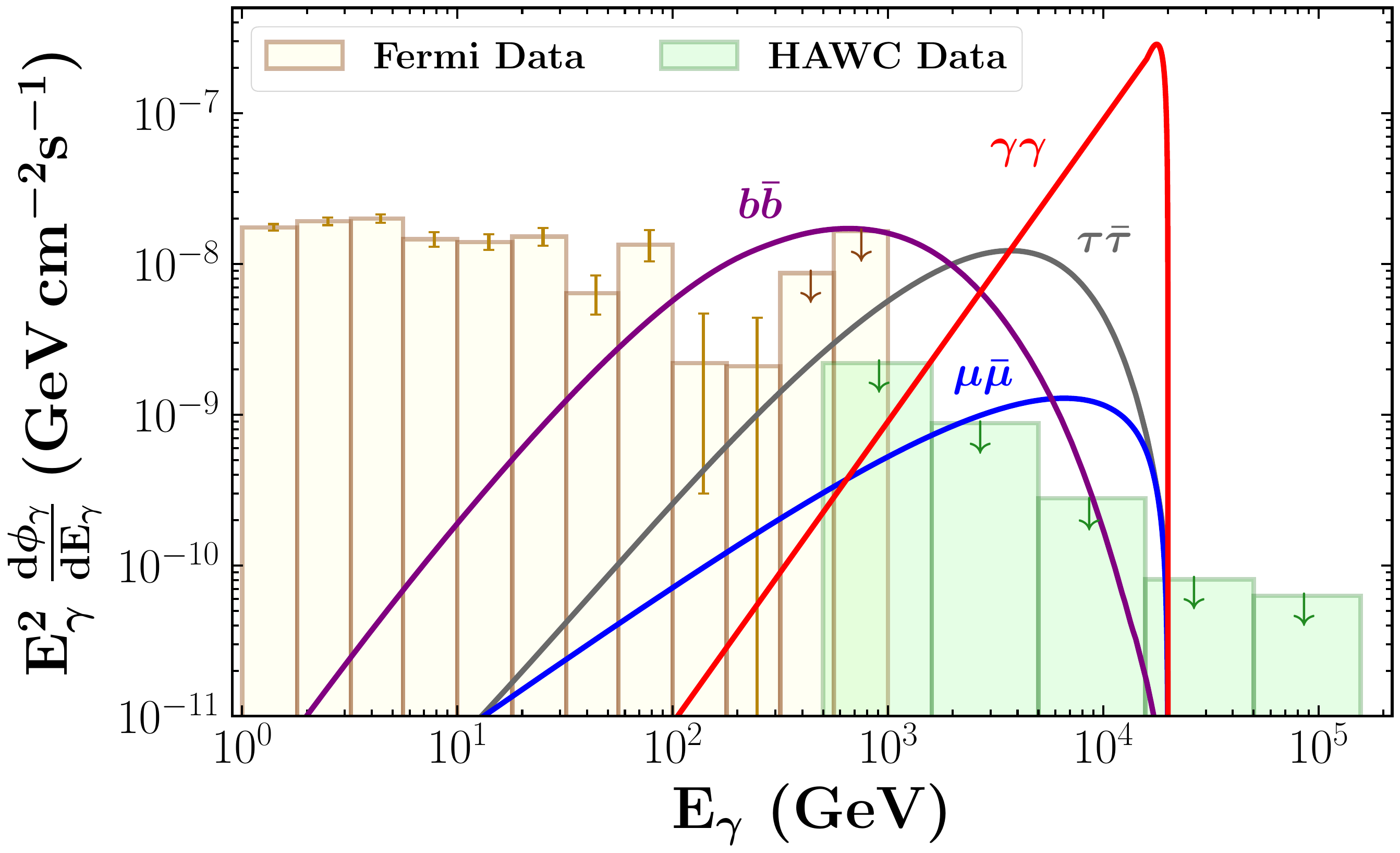}
\caption{Photon flux reaching at the surface of Earth for different SM final states: $\gamma \gamma$ (red), $\tau \bar{\tau}$ (gray), $b \bar{b}$ (purple) and $\mu \bar{\mu}$ (blue) for DM mass, $m_{\chi} = 20 \, {\rm TeV}$, mediator mass, $m_Y = 50 \, {\rm GeV}$ and DM-electron scattering cross-section, $\sigma_{\chi e} = 10^{-40} \, {\rm cm^2}$ are portrayed. We have also depicted the solar gamma ray flux measurements from Fermi-LAT \cite{Tang:2018wqp} and HAWC \cite{HAWC:2018rpf} with yellow and green histograms respectively.}
\label{fig:flux} 
\end{center}
\end{figure}

The number density of the captured DM inside the Sun is governed by the interplay of capture, annihilation, and evaporation. For heavier DM masses ($m_{\chi} \gtrsim 5 \, {\rm GeV}$), the evaporation becomes numerically insignificant \cite{Liang:2018cjn,Garani:2021feo}. We assume equilibrium between the capture and annihilation of DM in the Sun  that allows us to  relate the annihilation with the  capture rate as given below,
\begin{eqnarray}
\Gamma_{\rm ann} = \frac{1}{2} \, C_{\rm ann} \, N_{\chi}^2 = \frac{C_{\odot}}{2},
\end{eqnarray}
where $C_{\rm ann}$ is the coefficient of DM annihilation and $C_{\odot}$ denotes the capture rate defined in Eq. (\ref{eq:caprate}). This correlation makes our results independent of the annihilation cross-section. DM particles annihilate to SM states through mediator that has a sufficiently long lifetime ($\tau_Y$) or a large velocity boost ($\eta$) can escape from the solar environment ($L_Y = \eta c \tau_Y > R_{\odot}$). The mediator can decay to various SM final states which can give rise to observable photon flux at the Earth-based observatories \cite{Batell:2009zp,Leane:2017vag,Arina:2017sng,HAWC:2018szf,Niblaeus:2019gjk,Bell:2021pyy}. The differential photon flux at the surface of Earth originating from such mediator is given by
\begin{eqnarray}
E_{\gamma}^2 \frac{d \phi_{\gamma}}{d E_{\gamma}} = \frac{\Gamma_{\rm ann}}{4 \, \pi \, D_{\odot}^2} \times {\rm Br \left(Y \rightarrow SM \bar{SM} \right)} \times \left( E_{\gamma}^2 \, \frac{dN_{\gamma}}{dE_{\gamma}} \right) \times \left( e^{-\frac{R_{\odot}}{\eta c \tau_Y}} - e^{-\frac{D_{\odot}}{\eta c \tau_Y}} \right),
\label{eq:gamma_flux}
\end{eqnarray}
where ${\rm Br \left(Y \rightarrow SM \bar{SM} \right)}$ is the branching ratio to a given SM final state, $D_{\odot}$ is the distance between the Sun and the Earth. The gamma ray spectrum, $dN_{\gamma}/dE_{\gamma}$, is adopted from \cite{Elor:2015bho}. The last term in the parenthesis estimates the survival probability of the signal to reach terrestrial detectors. The survival probability is calculated assmuing a mediator of decay length $R_{\odot}$. Under equilibrium assumption, for a given decay route of the mediator, the photon flux for a specified DM and mediator mass is solely determined by the DM-electron scattering cross-section, $\sigma_{\chi e}$. Assuming $100 \%$ branching ratio, we present representative photon flux for various SM final states in Fig.\,\ref{fig:flux}. The mediator can in principle scatter with the SM constituents of the Sun owing to its SM coupling. The coupling and the boost parameter can be easily arranged so that the mediator can emerge from the Sun with minimal attenuation. For specific realization of this in particle physics models see \cite{Batell:2009zp,Feng:2016ijc,Smolinsky:2017fvb,Arina:2017sng,Bell:2021pyy}. While we do not conform to any specific models, we will assume that the attenuation is minimal and the mediator entirely produces decay signatures. The lifetime of the mediator have a lower bound in the nano-second range from collider experiments \cite{ATLAS:2020wjh} and upper limits from cosmological observations like BBN  is $\lesssim 1\,$s \cite{Kawasaki:2004yh,Depta:2020zbh}. These bounds are easily in consonance with the mediator considered here. Long-lived mediators can be realized in a large class of well-motivated models like the secluded DM \cite{Pospelov:2007mp,Pospelov:2008jd,Batell:2009zp,Dedes:2009bk,Fortes:2015qka,Okawa:2016wrr,Yamamoto:2017ypv}, dark photon \cite{Holdom:1985ag,Holdom:1986eq,Bell:2021pyy,Feng:2016ijc}, dark higgs \cite{Chen:2009ab} scenarios.

%%%%%%%%%%%%%%%%%%%%%%%%%%%%%%%%%%%%%%%%%%%%%%%%%%%%
%%%%%%%%%%%%%%%%%%%%%%%%%%%%%%%%%%%%%%%%%%%%%%%%%%%%

\section{Results}
\label{res}

State of the art solar gamma ray measurements in $0.1-10^3$ GeV energy range is provided by the Fermi-LAT satellite based experiment \cite{Fermi-LAT:2011nwz,Ng:2015gya,Tang:2018wqp}. The Fermi-LAT solar gamma ray flux measurements  \cite{Tang:2018wqp} are shown in Fig.\,\ref{fig:flux}. It's worth mentioning that the Fermi-LAT data in the last two bins are upper limits obtained from null measurements. The HAWC \cite{HAWC:2018rpf} and ARGO-YBJ \cite{ARGO-YBJ:2019mdq} have looked for solar gamma rays in the multi-TeV window, which complement the Fermi-LAT data. The more sensitive HAWC data, in the energy range $0.5$ TeV to $100$ TeV, has been reported in Ref. \cite{HAWC:2018rpf} and the corresponding $95 \%$ C.L. upper limits in gamma-ray flux is displayed in Fig.\,\ref{fig:flux}. The measurements of Fermi-LAT and HAWC are in the right ball park to constraint the photon signals from captured DM as can be seen from Fig.\,\ref{fig:flux}.

\begin{figure}[t]
\begin{center}
\includegraphics[scale=0.2]{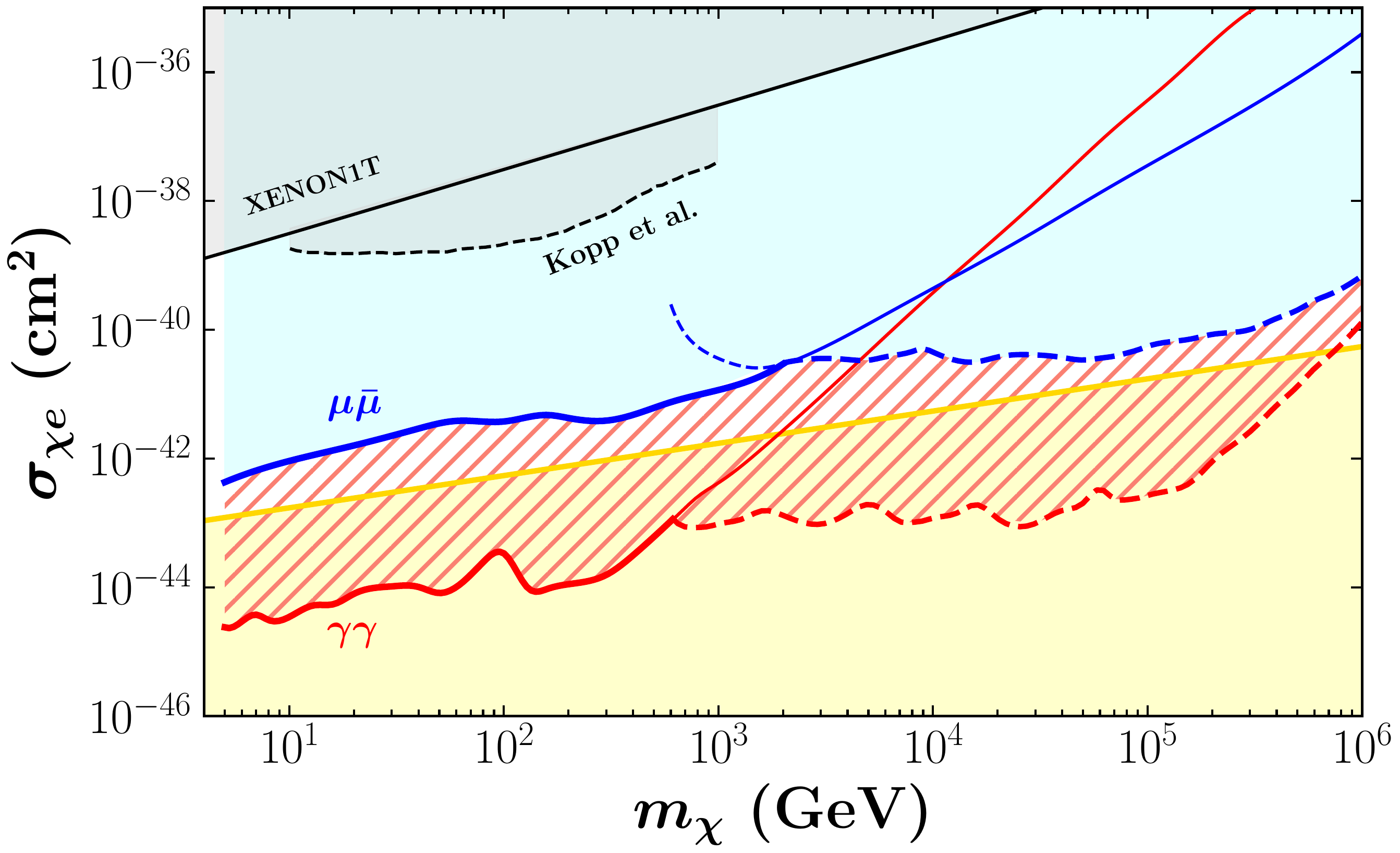}
\caption{Excluded regions of DM-electron scattering cross-section obtained from Fermi-LAT (HAWC) measurements have been shown for two different decay channels: $\gamma \gamma$ with red solid (dashed) line and above and $\mu \bar{\mu}$ with blue solid (dashed) line and above. We have also plotted the constraints obtained by looking at the direct annihilation of captured solar DM in Super-Kamiokande with black dashed line \cite{Kopp:2009et} as well as XENON1T bound with black solid line \cite{XENON:2019gfn,Emken:2021uzb}. The region where the equilibrium assumption does not hold for the typical solar age is represented with the yellow shading.}
  \label{fig:limits} 
\end{center}
\end{figure}

Gamma ray flux may originate from hadronic interactions of cosmic ray particles in the solar atmosphere \cite{Seckel:1991ffa,Zhou:2016ljf}. In addition, processes such as Inverse Compton scattering of cosmic-ray electrons with solar photons \cite{Orlando:2006zs,Moskalenko:2006ta,Orlando:2013pza} and particle acceleration during severe solar events \cite{Kafexhiu:2018wmh} can produce gamma ray flux indistinguishable  to the one investigated here. We keep the exclusion limits modest by assuming that the entire Fermi-LAT and HAWC observations are based on the photon flux from captured DM annihilation, neglecting the aforementioned backgrounds. Our limits will be stronger than the ones stated here if we incorporate all other processes as a background in the observed data. In Fig.\,\ref{fig:limits}, the excluded regions of DM-electron scattering cross-section from the Fermi-LAT and HAWC measurements, assuming $100 \, \%$ branching ratio to $\mu \bar{\mu}$ ($\gamma \gamma$) is represented by the blue (red) line and the region above it. The solid and dashed blue (red) contours represent the sensitivity arising from Fermi-LAT and HAWC observations respectively for $\mu \bar{\mu}$ ($\gamma \gamma$) final states. The discussed framework is limited by a minimum testable DM mass, owing to the fact that DM lighter than $\sim 5$ GeV would rapidly evaporate after getting captured within the Sun thus significantly reducing the signal \cite{Liang:2018cjn,Garani:2021feo}. The yellow shaded region depicts the breakdown of equilibrium where for typical  WIMP-like annihilation cross-section ($\langle \sigma v \rangle \sim 3 \times 10^{-26} \, {\rm cm^3 s^{-1}}$ \cite{Peter:2009mk, Steigman:2012nb, Saikawa:2020swg}) the captured DM do not equlibriate within the solar age \cite{Peter:2009mk}. In this regime the DM do not equlibriate within the Sun and consequently the limits within this region get considerably relaxed. Though for $\gamma \gamma$ channel, the limit penetrates into the non-equlibriation regime it can still rule out regions of the parameter space in the red hatched region. For mediator decay to $\mu \bar{\mu}$, the exclusion limits can reach upto $\mathcal{O}(10^{-43}-10^{-40}) \, \rm cm^2$ which keeps the capture in equilibrium with the annihilation within our parameter space of interest. Strikingly, for the considered scenario the present solar bounds on the DM-elctron couplings  from the Fermi-LAT and the HAWC pushes the excluded regions to the limit permitted by the equilibrium floor, as can be seen in Fig.\,\ref{fig:limits}. The limit obtained by probing neutrino signal of captured DM in Sun is shown for reference \cite{Kopp:2009et}. For XENON1T, we obtain the limits by considering S2-only analysis utilizing the \texttt{obscura} code \cite{XENON:2019gfn,Emken:2021uzb}. Note that our limits mildly depend on the mediator mass which is set at 5 GeV. The bounds obtained from this analysis, though model dependent, provide the most stringent bounds on the DM-electron scattering cross-section in the considered region of DM parameter space improving the existing bounds by a factor of $\sim 10^4-10^6$.

%%%%%%%%%%%%%%%%%%%%%%%%%%%%%%%%%%%%%%%%%%%%%%%%%%%%
%%%%%%%%%%%%%%%%%%%%%%%%%%%%%%%%%%%%%%%%%%%%%%%%%%%%

%%%%%%%%%%%%%%%%%%%%%%%%%%%%%%%%%%%%%%%%%%%%%%%%%%%%
%%%%%%%%%%%%%%%%%%%%%%%%%%%%%%%%%%%%%%%%%%%%%%%%%%%%

\section{Conclusions}
\label{con}

Sun, being our host star, is an excellent celestial laboratory to probe non-gravitational interactions of DM. The infalling DM can scatter off electrons inside the Sun and be captured within the solar interior. The annihilation of these captured DM through long-lived mediators can produce considerable gamma-ray flux that can be observed in terrestrial detectors. In this work, we have explored the possibility of searching for the annihilation signatures of electrophilic DM captured inside the Sun utilizing the Fermi-LAT and the HAWC data. We obtain conservative limits by comparing the gamma-ray flux from captured DM annihilating through long-lived mediators with the solar disk measurements of the gamma ray at the Fermi-LAT and the HAWC. We find that in our parameter space of interest, the limits are orders of magnitude stronger than the existing bounds. Depending on the mediator decay mode, the current sensitivity of the observational data approaches the Sun's equilibrium floor, effectively covering the parameter space that can be explored within this framework.

%%%%%%%%%%%%%%%%%%%%%%%%%%%%%%%%%%%%%%%%%%%%%%%%%%%%
%%%%%%%%%%%%%%%%%%%%%%%%%%%%%%%%%%%%%%%%%%%%%%%%%%%%

%%%%%%%%%%%%%%%%%%%%%%%%%%%%%%%%%%%%%%%%%%%%%%%%%%%%
\paragraph*{Acknowledgments\,:} We thank Biplob Bhattacherjee for discussions and Sambo Sarkar for help with the plots. DB acknowledges MHRD, Government of India for fellowship. TNM acknowledges IOE-IISc fellowship program for financial assistance. TSR acknowledges ICTP, Trieste for hospitality under the Associateship Program during the completion of this work.
%%%%%%%%%%%%%%%%%%%%%%%%%%%%%%%%%%%%%%%%%%%%%%%%%%%%

\bibliographystyle{JHEP}
\bibliography{capelec_ref.bib}
\end{document}